\documentclass[preprint,12pt]{elsarticle}
\usepackage{graphicx}
\usepackage{subfigure}
\usepackage{multirow}
\usepackage{booktabs}
\usepackage{rotating}
\usepackage{natbib}
\setcitestyle{authoryear,open=(,close=)}
\usepackage{bm,amssymb,amsthm,amsmath}
\newcommand{\nc}{\newcommand}
\nc{\mbb}{\mathbb}\nc{\bb}{\mathbb}
\nc{\mbf}{\mathbf}\nc{\mb}{\mathbf}
\nc{\mc}{\mathcal}
\nc{\matr}[1]{\bm{#1}}

\usepackage{algorithm,algorithmic}

\journal{Spatial Statistics}

\begin{document}

\begin{frontmatter}

\title{Approximately Optimal Spatial Design: How Good is it?}

%% include affiliations in footnotes:
\author[firstaddress]{Yu Wang}

\author[secondaddress]{Nhu D. Le}

\author[thirdaddress]{James V. Zidek\corref{correspondingauthor}}
\cortext[correspondingauthor]{Corresponding author}
\ead{jim@stat.ubc.ca}

\address[firstaddress]{University of Michigan, 500 S State St, Ann Arbor, MI USA}
\address[secondaddress]{BC Cancer Research Centre, 675 W 10th Ave, Vancouver, BC Canada}
\address[thirdaddress]{University of British Columbia, 2329 West Mall, Vancouver, BC Canada}

\begin{abstract}
The increasing recognition of the association between adverse human health conditions and many environmental substances as well as processes has led to the need to monitor them. An important problem that arises in environmental statistics is the design of the locations of the monitoring stations for those environmental processes of interest. One particular design criterion for monitoring networks that tries to reduce the uncertainty about predictions of unseen processes is called the maximum-entropy design. However, this design criterion involves a hard optimization problem that is computationally intractable for large data sets. Previous work of \citet{2017arXiv170900151W} examined a probabilistic model that can be implemented efficiently to approximate the underlying optimization problem. In this paper, we attempt to establish statistically sound tools for assessing the quality of the approximations. 
\end{abstract}

% \begin{highlights}
% \begin{itemize}
%     \item Research highlight 1
%     \item Research highlight 2
% \end{itemize}
% \end{highlights}


\end{frontmatter}

\section{Introduction}\label{intro}
The design of experiments in classical statistics addresses the problem of making inferences about a set of random variables given observations of just a subset of them.   There an experimenter selects and runs a well planned set of experiments to optimize a process or system from well supported conclusions about the  behaviour of that process of system. In environmental statistics, the experiment yields observations of a certain environmental process (temperature, air pollution, rainfall, etc) taken from a set of monitoring stations. Since usually maintaining all stations would be costly and hence infeasible, one may need to select only a subset of them. 

But how should that subset be chosen?  Formally, given a set $ N $ of $ n $ points, called the design space, that would mean choosing a design size $ k \leq n $  and then a subset $ K \subset N $ of $ k $ points. Most approaches to spatial design fall into the following categories~\citep{zidek2019monitoring}:
\begin{itemize}
    \item Geometry-based: It involves heuristic arguments and includes such things as regular lattices, triangular networks, or space-filling designs~\citep{cox1997spatial,royle1998algorithm,nychka1998design}. The heuristics may reflect prior knowledge about the environmental process of interest. These designs can be especially useful when the design's purpose is exploratory~\citep{muller2005comparison}.
    \item Probability-based: This approach to design has been used widely not only in environmental contexts, but also for a variety of other purposes such as public opinion polling and nation-wide surveys. It has the obvious appeal that sample selection is based on the technique of sampling at random from a list of the population elements (the sampling frame). Thus in principle (though not in practice) the designers need not have any knowledge of the population of interest. Moreover, they may see competing design methods as biased because they rely on prior knowledge of the population, usually expressed through statistical models, which involve assumptions about the nature of the process under investigation, of which can be wrong. Thus, those methods may skew the selection and biasing the inference about the process being monitored. 
    \item Model-based: The majority of designs for environmental monitoring networks rely on the model-based approaches. Broadly speaking, model-based designs optimize some form of inference about the process or its model parameters. Although the models do indeed skew the selection process, they do so in accord with prior knowledge and can make the design maximally efficient in extracting relevant information for inferences about the process. In contrast the probability-based approach may be seen as gambling on the outcome of the randomization procedure and hence risking the possibility of getting designs that ignore aspects of the process that are important for inference.
\end{itemize}

In this paper, we restrict our discussion to the model-based design. Given a well--defined optimality criterion, this task would be relatively straightforward in principle at least. Thus monitoring networks may be set up for collecting data: near a point source such as a smelter to determine emissions levels; to optimally estimate parameters in a spatial regression model; to determine the impact of the startup of a
new source of emissions.

However networks set up for one purpose may in time used for another. For example data collected to detect noncompliance with air quality standards are routinely used to study the impact of air pollution on human health. Networks set up to measure acidic precipitation were later appropriated for use in air quality monitoring programs. The latter exemplifies a situation where the current use of the network was not even foreseen by the designers. In short networks may end up with a multiplicity of objectives and even some that were not foreseen when they were being created. In other words, no well-defined criterion need exist when networks are created. In that case a reasonable surrogate would call for the $ k $ points to have the greatest uncertainty with respect to their joint distribution. For then the measurement of their responses would maximally reduce the uncertainty about the process. That in turn suggests maximizing the joint relative entropy of the selected points since their measurement would eliminate that uncertainty altogether. Then since the total entropy for all the points in $ N $ is fixed, an entropy decomposition theorem~\citep{caselton1984optimal} implies that that design would  also reduce the joint uncertainty about the predictor of the unmeasured responses based on the data from those in the design set. Simultaneously the joint uncertainty about the parameters of the prediction model would also be minimized. 

The problem of computing the entropy of a proposed design is simplified for Gaussian and Student-t processes, for then the entropy is a simple transformation of a symmetric positive definite $n \times n$ matrix $\matr{L}$ indexed by the set $N$, in fact a covariance matrix. Then the entropy associated with any $k$--element subset $K$ of $N$, up to a known positive affine transformation, is the logarithm of the determinant of the $k \times k$ principal submatrix $\matr{L}[K]$ with row and column indices in $K$ (see~\citet{caselton1984optimal} for details).

However, as demonstrated in~\citet{ko1995exact}, this optimization problem is NP--hard. \citet{2017arXiv170900151W} proposed an approximation strategy to this combinatorial optimization problem based on the determinantal point process (DPP). This novel approximation algorithm is stochastic, unlike other existing methods in the literature, and always approaches the optimum as the number of iterations increases. The proposed algorithm can easily be parallelized; thus multiple computer processing units could be used simultaneously to increase computing power. As shown in~\citet{2017arXiv170900151W}, the algorithm is computationally efficient as measured by its running time. In this paper, we further investigate the DPP approximations and show that they can be considered as record values, and hence theoretical tools from record values theory can be used to study the behavior of those approximations. 

The remainder of the paper is organized as follows. In Section~\ref{overview}, we formally define the problem and give an overview of existing algorithms for finding/approximating the optimal design, emphasizing on the DPP algorithm. In Section~\ref{records}, we connect DPP approximations with record values and show that tools from record-value theory can be used to analyze the quality of the approximations. We conclude in Section~\ref{conclusion} and comment on challenges and possible directions in future research of spatial design.

\section{Overview of Algorithms for Finding Optimal Designs}\label{overview}
\subsection{Definitions and notation}
Let $N = \{1, 2, 3, ..., n\}$ where $n$ is a positive integer. We use $\matr{K}$ to denote a real symmetric positive definite matrix indexed by elements in $N$. Further, let $S$ be an $s$--element subset of $N$ with $1 \leq s \leq n$. Let $\matr{K}[S, S]$ denote the principal submatrix of $\matr{K}$ having rows and columns indexed by elements in $S$--note that $\matr{K}[S, S] = \matr{K}[S]$. Write $v_{N} (S) = \text{det}(\matr{K}[S])$ to denote the determinant of the matrix $\matr{K}[S]$. Our optimization problem is to determine
\begin{equation}\label{eqn:P}
	\max_{{S}: |S|=s, S \subset N} v_{N} (S),
\end{equation} and the associated maximizer $S$. 

\subsection{Finding a solution}
Numerous algorithms have been developed for solving/approximating the optimization problem, including both exact methods and heuristics. For small, tractable problems (e.g., up to size $\binom{50}{25}$), efficient software implementation of complete enumeration, such as that in \texttt{EnviroStat v0.4-0 R} package~\citep{EnviroStat} works reasonably well. \citet{ko1995exact} first introduced a branch--and--bound algorithm that guarantees an optimal solution. Specifically, the authors established a spectral upper bound for the optimum value and incorporated it in a branch--and--bound algorithm for the exact solution of the problem. Although there have been several further improvements, mostly based on incorporating different bounding methods~\citep{anstreicher1996continuous,anstreicher1999using,hoffman2001new,lee2000semidefinite,lee2003linear}, the algorithm still suffers from scalability challenges and can handle problem of size only up to about $n=75$. Most recently, \citet{anstreicher2018linx} introduced a new bound ``linx" based on a simple but previously unexploited determinant identity. With linx, the branch--and--bound algorithm can solve some instances of the problem of size $n=124$.    

\subsubsection{Greedy algorithm} 
For large intractable problems,  heuristics all lacking some degree of generality and theoretical guarantees on achieving proximity to the optimum, can be used to find reasonably good solutions. One of the best known is the DETMAX algorithm of~\citet{mitchell1974algorithm}, based on the idea of exchanges, which is widely used by statisticians for finding approximate $D$--optimal designs. Due to the lack of readily available alternatives, \citet{Guttorp1992UsingEI} use a greedy approach, which is summarized in Algorithm~\ref{alg:GREEDY}. \citet{ko1995exact} experiment with a backward version of the Algorithm~\ref{alg:GREEDY}: start with $S = N$, then, for $j = 1, 2, ..., n-s$, choose $l \in S$ so as to maximize $v_{N} (S \setminus \{l\})$, and then remove $l$ from $S$. They also describe an exchange method, which begins from the output set $S$ of the greedy algorithm, and while possible, choose $k \in N \setminus S$ and $l \in S$ so that $v_{N} (S \cup \{k\} \setminus \{l\}) > v_{N} (S)$, and replace $S$ with $S \cup \{k\} \setminus \{l\}$.

\begin{algorithm}[tbh]
\caption{Greedy Algorithm}
\label{alg:GREEDY}
\begin{algorithmic}
\REQUIRE Size $k$ and an empty set $S=\emptyset$.
\FOR{$i = 1, \dots, k$}
        \STATE Choose $s \in N \setminus S$ so as to maximize $v_{N} (S \cup \{s\})$.
        \STATE Set $S=S \cup \{s\}$.
\ENDFOR
\ENSURE Set $S$ with $k$ elements.
\end{algorithmic}
\end{algorithm}

\subsubsection{Genetic algorithm}
More recently, \citet{ruiz2010stochastic} propose a stochastic search procedure based on Genetic Algorithm (GA)~\citep{holland1975adaptation} for finding approximate optimal designs for environmental monitoring networks. They test the algorithm on a set of simulated datasets of different sizes, as well as on a real application involving the redesign of a large--scale environmental monitoring network. In general, the GA seeks to improve a population of possible solutions using principles of genetic evolution such as natural selection, crossover, and mutation. The GA considered here consists of general steps described in Algorithm \ref{alg:GA}. The GA has been known to work well for optimizing hard, black--box functions with potentially many local optima, although its solution is fairly sensitive to the tuning parameters~\citep{goldberg1988genetic,whitley1994genetic}.

\begin{algorithm}[tbh]
\caption{Genetic Algorithm}
\label{alg:GA}
\begin{algorithmic}[1]
%\REQUIRE Size $k$ and the kernel matrix $\matr{L}$.
\STATE Choose at random an initial population of size $N_0$, that is, a set of $N_0$ possible solutions $S_1,...,S_{N_0}$.
\STATE Compute the fitness, that is, the value of the objective function $v_{N}(S_i)$, $i = 1,...,N_0$, for each of the solutions in the population.
\STATE \textit{Crossover}: choose a proportion, $p_{\text{cross}}$, of solutions from the population. These solutions are selected according to a fitness-dependent selection scheme. Among these selected solutions, pairs of solutions are formed at random.
\STATE \textit{Mutation}: choose a proportion, $p_{\text{mutprop}}$, of solutions from the population with equal probability. For each selected solution, each gauged site may be swapped, according to a mutation probability $p_{\text{mut}}$, with a randomly chosen ungauged neighbour site.
\STATE Compute the fitness of the solutions obtained by crossover and mutation. Include these solutions in the current population, creating an augmented population.
\STATE \textit{Selection}: the population of solutions of the new generation will be selected from this augmented population. A proportion of solutions with best fitness, called elite, enter directly in the new generation while the remaining members of the new generation are randomly chosen according to certain fitness--dependent selection scheme (see \citet{goldberg1991comparative} for a discussion of various selection schemes).
\STATE Stop the algorithm if the stop criterion is met. Otherwise, return to Step 3.
%\ENSURE The maximum determinant and the associated set of indices.
\end{algorithmic}
\end{algorithm}

\subsection{Determinantal point processes}\label{dpp}
\citet{2017arXiv170900151W} develop an efficient stochastic search algorithm based on the determinantal point process. We briefly review the basics of the DPP and the stochastic search algorithm. Recall that a point process $\mathbb{P}$ on the ground set $G = \{1, 2, ..., n\}$ is a probability measure defined on the power set of $G$, i.e., $2^{G}$. A point process $\mathbb{P}$ is called a determinantal point process, if when $Y$ is a random subset drawn according to $\mathbb{P}$, then we have for every $S \subseteq Y$,
\begin{equation}\label{marginal}
	\mathbb{P}(S \subseteq Y) = \text{det}(\matr{K}[S]),
\end{equation} for some matrix $\matr{K} \in \mathbb{R}^{n \times n}$
indexed by the elements of $G$ that is symmetric, real and positive semidefinite, and satisfies $0 \leq \bm{a}^T \matr{K} \bm{a} \leq 1$ for any $\bm{a} \in \mathbb{R}^{n \times 1}$.

In practice, it is more convenient to characterize DPPs via $L$--ensembles~\citep{borodin2005harmonic,kulesza2012determinantal}, which directly define the probability of observing each subset of $G$. An $L$--ensemble defines a DPP through a real positive semidefinite matrix $\matr{L}$, indexed by the elements of $G$, such that
\begin{equation}\label{direct}
	\mathbb{P}_{\matr{L}}(\mathbf{Y} = Y) = \frac{\text{det}(\matr{L}[Y])}{\sum_{Y^\prime \subseteq G} \text{det}(\matr{L}[Y^\prime])},
\end{equation} where the normalizing constant $\sum_{Y^\prime \subseteq G} \text{det}(\matr{L}[Y^\prime]) = \text{det}(\matr{L} + \matr{I})$ and $I$ is an $n \times n$ identity matrix. Equation \eqref{direct} represents the probability of exactly observing all possible realizations of $\mathbf{Y}$. 

Note that standard DPP models described above may yield subsets of any random size. A $k$--DPP on a discrete set $G = \{1, ..., n\}$ is simply a DPP with fixed cardinality $k$. It can be obtained by conditioning a standard DPP on the event that the set $Y$ has cardinality $k$, as follows
\begin{equation}\label{kdpp}
	\mathbb{P}_{\matr{L}}^k (Y) = \mathbb{P}(\mathbf{Y}=Y | |Y|=k) = \frac{\text{det}(\matr{L}[Y])}{\sum_{|Y^{\prime}|=k} \text{det}(\matr{L}[Y^{\prime}])},
\end{equation} where $|Y|$ denotes the cardinality of $Y$. This notion is essential in the context of our cardinality--constrained discrete optimization problem. 

The sampling of a $k$--DPP largely relies on being able to express DPP as a mixture of elementary DPPs~\citep{kulesza2012determinantal}, also commonly known as determinantal projection processes. Using Algorithm \ref{alg:kDPPsample} as adapted from~\citet{kulesza2012determinantal}, the sampling from a $k$--DPP can be performed in $\mathcal{O}(N^3)$ time in general, and every $k$--element subset $S$ among the $n$ candidate points has the opportunity to be sampled with probability given in Equation \eqref{kdpp}.

To handle the NP--hard optimization problem~\eqref{eqn:P}, the $k$--DPP sampling approach involves generating such $k$--DPP subsets repeatedly and calculating the objective function $v_{N} (S)$, such that successively better approximations, as measured by $v_{N} (S)$, can be found. The approximate solution to \eqref{eqn:P} is then given by the best $v_{N} (S)$ attained up to a certain number of simulations and its associated indices of points, as described in Algorithm \ref{alg:SamplingSoln}. 
Note that eigendecomposition of the kernel matrix can be done as a pre--processing step and therefore does not need to be performed before each sampling step. Therefore, assuming that we have an eigendecomposition of the kernel in advance, sampling one $k$--DPP run in $\mathcal{O}(Nk^3)$ time~\citep{kulesza2012determinantal}, and the computation of the determinant of a submatrix typically takes $\mathcal{O}(k^3)$ time. Overall, Algorithm \ref{alg:SamplingSoln} runs in $\mathcal{O}(Nk^3)$ time per iteration. 

\begin{algorithm}[tbh]
\caption{Sampling from a $k$--DPP}
\label{alg:kDPPsample}
\begin{algorithmic}
\REQUIRE size $k$ and $\{{\bf v}_n, \lambda_n\}$ eigenvectors and eigenvalues of $\matr{L}$.
\STATE $J \leftarrow \emptyset$.
\STATE Compute elementary DPPs $E_{1}^{n}, \dots, E_{k}^{n}$, for $n = 0, \dots, N$.
\FOR{$n = N, \dots, 1$}
        \STATE Sample $u \sim U[0,1]$
        \IF{ $u < \frac{\lambda_n E_{k-1}^{n-1}}{E_{k}^{n}}$}
        \STATE $J \leftarrow J \cup \{n\}$
        \STATE $k \leftarrow k-1$
        \IF{k = 0}
        \STATE {\bf break}
        \ENDIF
        \ENDIF
\ENDFOR
\STATE $V \leftarrow \{{\bf v}_n\}_{n \in J}$
\STATE $Y \leftarrow \emptyset$
\WHILE{$|V| > 0$}
\STATE Select $y_i$ from $Y$ with probability given by $\frac{1}{|V|} \sum_{{\bf v} \in V} ({\bf v}^{\top} {\bf e}_i)^2$
\STATE $Y \leftarrow Y \cup \{y_i\}$
\STATE $V \leftarrow V_{\bot}$, an orthonormal basis for the subspace of $V$ orthogonal to ${\bf e}_i$
\ENDWHILE
\ENSURE $Y$.
\end{algorithmic}
\end{algorithm}
\begin{algorithm}[tbh]
\caption{Sampling--based solution strategy using $k$--DPP}
\label{alg:SamplingSoln}
\begin{algorithmic}[1]
\REQUIRE Size $k$ and the kernel matrix $\matr{L}$.
\STATE Sample $k$ indices according to the $k$-DPP distribution specified by $\matr{L}$ using Algorithm \ref{alg:kDPPsample}.
\STATE Compute determinant of the submatrix indexed by the $k$ indices sampled.
\STATE Repeat Step $1$ and $2$ until the maximum number of iterations or the maximum computing resources.
\ENSURE The maximum determinant and the associated set of indices.
\end{algorithmic}
\end{algorithm}

\subsection{Illustrative application: optimizing maximum-entropy designs for monitoring networks}\label{dppexp}
For numerical illustration we consider the data supplied by the U.S. Global Historical Climatology Network--Daily (GHCND), which is an integrated database of climate summaries from land surface stations across the globe. For illustrative purposes, we selected $97$ temperature monitoring stations where the maximum daily temperature was recorded. A subset of $67$ stations was selected among the $97$ stations to constitute a hypothetical monitoring network. An additional $30$ stations were selected and designated as potential sites for new monitors. \citet{casquilho2018design} has successfully approximated a maximum-entropy design using $k$-DPP for an instance of this problem. In this case study, the goal is to select a subset of $10$ stations from among the additional $30$ to augment the network based on the maximum-entropy design criterion. 

Using the notation presented in Equation~\eqref{eqn:P}, $\matr{C}$ here is the estimated covariance matrix of $30$ candidate sites and $S$ is the subset of $10$ sites that maximize $H_N(S)$. For tractable optimization problems, the maximal value of the objective function (or equivalently the optimal design) can be obtained efficiently by the branch-and-bound algorithm detailed in~\citet{ko1995exact}. In particular, the maximal value is $80.09011$.

\begin{figure}[h]
    \centering
	\includegraphics[width=\textwidth]{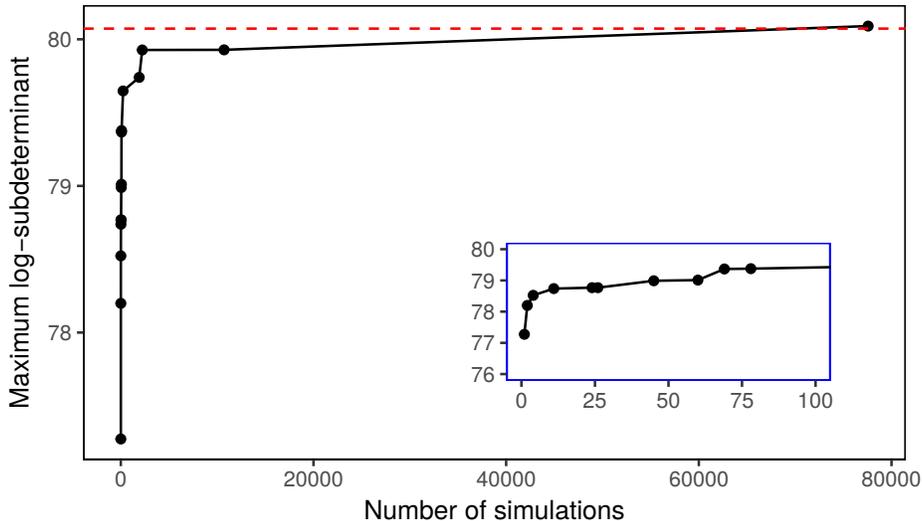}
	\caption{Occurrence of maximum log--determinants of the restricted conditional hypercovariance matrix when increasing the number of simulated $10$-DPP samples. The optimum solution is marked by the red horizontal dashed red line. The inset shows a zoomed-in view of the first $100$ samples.}
    \label{fig:dppreal}
\end{figure}

For comparison, we first performed the greedy algorithm discussed in Algorithm~\ref{alg:GREEDY}, which yielded a solution of $80.07284$. Using Algorithm~\ref{alg:GA} with the tuning parameters suggested in \citet{ruiz2010stochastic} ($N_0 = 100$, $p_{\text{cross}} = 0.75$, $p_{\text{mut}} = 0.05$, and a tournament selection scheme with four competitors), the GA yielded a solution of $80.09011$ after $1000$ generations. Similarly, the proposed $10$-DPP achieved the optimal value after about $80,000$ simulations. As illustrated in Figure~\ref{fig:dppreal}, the maximal value of the log--determinant among the simulations increases as the number of simulations escalates.  

In terms of computation time, for this particular problem about $20$ minutes of wall clock time was taken to simulate $100,000$ subsets from the $10$-DPP using \texttt{R} programming language~\citep{R} on a laptop with a \texttt{2.5 GHz Intel Core i7} processor and a \texttt{16 GB 1,600 MHz DDR3} RAM. In the same computational environment, $5$ minutes of wall clock time was needed to simulate $1,000$ generations of GA; less than $1$ second of wall clock time was needed for the greedy algorithm implemented in the \texttt{edesign} package in \texttt{R} to yield a solution.

\section{$k$-DPP Approximations as Record Values}\label{records}

\subsection{Definitions and notation}
The standard record value process corresponding to an infinite sequence of i.i.d. observations is our focus. Let $X_1,X_2,\dots$ be an infinite sequence of random variables having the same distribution as the random variable $X$. Denote the cumulative distribution function (cdf) of $X$ by $F$. Usually an assumption that $F$ is continuous is invoked to avoid the possibility of ties and allow a cleaner theoretical development. 

An observation $X_j$ is called an upper record value (or simply a record) if its value exceeds that of all previous observations, that is, $X_j > X_i$ for all $i<j$. The times at which records appear are often of interest. For simplicity, let us assume that $X_j$ is observed at time $j$. Then the sequence of record time $\{T_n,n \geq 0\}$ is defined as follows:
\begin{equation}
	T_n = \min \{j:X_j > X_{T_{n-1}}\}
\end{equation} for $n \geq 1$ and $T_0=1$ with probability $1$. The sequence of record values $\{R_n\}$ is then defined by
\begin{equation}
	R_n = X_{T_n}, \quad n=0,1,2,\dots \ .
\end{equation} Since the first observation in any sequence will always be a record, here $R_0$ is sometimes referred to as the reference value or trivial record. 

% The above definition of record sequence implicitly assumes that the cdf $F$ will not produce any unbreakable record. This will not be the case if there exists some value $x_0$ such that $F(x_0)-F(x_0-)>0$ and $F(x_0)=1$, that is, if there is a largest possible real value that can be obtained with positive probability. Though such cases are sometime of interest and in fact, this is the case in our study, we will eliminate them from our discussion for now and assume that the cdf $F$ cannot yield unbreakable records so that $\{R_n,n \geq 0\}$ will be a strictly increasing non-terminating sequence (we will explain in later sections why we still can apply such record values theory to our study). Note that this definition does not imply that $\{R_n\}$ is unbounded. For example, if the cdf $F$ is uniform on the interval $(0,1)$, then $\{R_n\}$ satisfies the definition and is bounded above by $1$.

We may also define a record increment (jump) sequence $\{J_n,n \geq 0\}$ by
\begin{equation}
	J_n = R_n - R_{n-1}
\end{equation} for $n \geq 1$ and $J_0=R_0$. An inter-record time sequence $\Delta_n$, is also of interest and can be defined as
\begin{equation}
	\Delta_n = T_n - T_{n-1}, \quad n=1,2,\dots \ .
\end{equation} Finally, the number of records may be tracked by a counting process $\{N_n,n \geq 1\}$, where
\begin{equation}
	N_n = \{\text{number of records among} \ X_1,\dots,X_n\}.
\end{equation}

As we shall see, in the setup of the classical record model defined above, where $X_i's$ are i.i.d. observations from a continuous distribution $F$, the record counting statistics $T_n$, $\Delta_n$, and $N_n$ will not be affected by $F$, unlike the records that are.

\subsection{Basic distributional results for record values}
In this section, we discuss distributional results for record values and related statistics (i.e., $R_n$, $T_n$, $\Delta_n$, $N_n$) from the classical model. It turns out that, for the classical model, strong arguments can be made in favour of studying i.i.d. exponentially distributed $X_i's$. In fact, the relationship between a standard exponential distribution and a general continuous distribution can be built to derive distributional results for record values from a general continuous cdf $F$. \citet{chandler1952distribution} first derived the survival function of the $n$th record corresponding to an i.i.d. sequence of random variables with cdf $F$:
\begin{equation}\label{eqn:recordsCDF}
	P(R_n > r) = (1-F(r))\sum_{k=0}^{n}\frac{(-\log(1-F(r)))^k}{k!},
\end{equation} or equivalently, the incomplete Gamma function:
\begin{equation}
	P(R_n \leq r) = \frac{1}{n!}\int_0^{-\log(1-F(r)} v^n e^{-v} dv.
\end{equation}
If $F$ is absolutely continuous with probability density function (pdf) $f$, we may differentiate Equation~\eqref{eqn:recordsCDF} to obtain the pdf for $R_n$:
\begin{equation}\label{eqn:recordsPDF}
	f_{R_n}(r) = f(r)\frac{(-\log(1-F(r)))^n}{n!}.
\end{equation} We may also obtain the joint pdf of the set of records $R_0,R_1,\dots,R_n$:
\begin{align}
	f_{R_0,R_1,\dots,R_n}(r_0,r_1,\dots,r_n) &= \frac{\prod_{i=0}^n f(r_i)}{\prod_{i=0}^{n-1}(1-F(r_i))} \\
	&= f(r_n) \prod_{i=0}^{n-1} h(r_i),
\end{align} where $h(r)=\frac{dH(r)}{dr}=\frac{f(r)}{1-F(r)}$ represents the hazard function. 

\subsection{Record times and related statistics}
As mentioned earlier, under the assumption that $F$, the common cdf of $X_i's$, is continuous, the distribution of record times and counts ($T_n's$,$N_n's$,$\Delta_n's$) does not depend on $F$. In order to discuss their distributional properties, we introduce a sequence of record indicator random variables as follows:
\begin{equation}
	I_n = I(X_n > \max\{X_1,\dots,X_{n-1}\}),
\end{equation} for $n>1$ and $I_1=1$ with probability $1$. It is not difficult to verify that the $I_n's$ are independent variables  with
\begin{equation}\label{eqn:recordIndicator}
	P(I_n=1) = \frac{1}{n}, \quad n \geq 1.
\end{equation} In other words, $I_n$ is a Bernoulli random variable with success probability $p=\frac{1}{n}$ with $n \geq 1$. It is more convenient to give an intuitive interpretation to this result: after the first record (which is the first observation in a sequence with probability $1$) the second observation has a probability of $\frac{1}{2}$ of beating the first record; the third observation could be either smaller than the first observation, between the first and the second observation, or larger than the second observation, which give it a probability of $\frac{1}{3}$ for beating the previous record. In general, each random variable has the same chance of being the largest and hence, the $n$th observation has a probability of $\frac{1}{n}$ of being a record. With this reasoning, the results given by \eqref{eqn:recordIndicator} easily follows.

With the help of $I_n$, we have the following facts about the record counting process $\{N_n,n \geq 1\}$:
\begin{equation}
	N_n = \sum_{i=1}^n I_i.
\end{equation} Since $I_i's$ are independent Bernoulli random variables, we immediately see that
\begin{equation}
	\mbb{E}N_n = \sum_{i=1}^n \frac{1}{i} \approx \ln n
\end{equation} and
\begin{equation}
	\text{Var}N_n = \sum_{i=1}^n \frac{1}{i}(1-\frac{1}{i}) \approx \ln n.
\end{equation} As we just confirmed, records are not common. In a sequence of $1000$ observations we expect to see only about $7$ records. Another fact that immediately follows is that the expected number of records goes to infinity as the number of samples gose to infinity, which is due to the divergence of harmonic series. In fact, we also have that $N_n \rightarrow \infty$ as $n \rightarrow \infty$ (see \citet{glick1978breaking} for a proof). The precise distribution of $N_n$ is complicated, but the expressions have been given by several authors~\citep{renyi1962theorie,david1962combinatorial,karlin1966first}:
\begin{equation}
	P(N_n = k) = \frac{S_n^{(k)}}{n!} \approx \frac{\ln(n)^k}{nk!}
\end{equation} for large sample size $n$, where $S_n^{(k)}$ is the Stirling number of the first kind.

We may use the information regarding the distribution of $N_n$ and $I_n$ to discuss the distribution of the $k$th non-trivial record time $T_k$. Note that the events $\{T_k=n\}$ and $\{I_n=1,N_{n-1}=k\}$ are equivalent. Consequently, 
\begin{equation}
	P(T_k = n) = P(I_n=1,N_{n-1}=k).
\end{equation} Since the events $\{I_n=1\}$ and $\{N_{n-1}=k\}$ are independent, we have
\begin{equation}
	P(T_k=n) = \frac{1}{n}\frac{S_{n-1}^{(k)}}{(n-1)!} = \frac{S_{n-1}^{(k)}}{n!}.
\end{equation} Note that $T_k$ grows rapidly as $k$ increases. In fact, \citet{glick1978breaking} verified that 
\begin{equation}
	\mbb{E}T_k = \infty, \quad \forall k \geq 1
\end{equation} and
\begin{equation}
	\mbb{E}\Delta_k = \mbb{E}(T_k-T_{k-1}) = \infty, \quad \forall k \geq 1.
\end{equation}

\citet{neuts1967waitingtimes} first developed an approach for computing the exact distribution of $\Delta_k$. By conditioning on the value of $R_{k-1}^{\ast}$ (assuming for convenience without loss of generality that the $X's$ are standard exponential random variable), we have, for $j=1,2,.\dots$, $k=1,2,\dots$
\begin{equation}
	P(\Delta_k > j) = \int_0^{\infty} \frac{x^{k-2}}{\Gamma(k)} e^{-x}(1-e^{-x})^j dx.
\end{equation} The integration is readily performed if we expand the term $(1-e^{-x})^j$, which yields
\begin{equation}
	P(\Delta_k > j) = \sum_{m=0}^j \binom{j}{m}(-1)^m \frac{1}{(1+m)^k},
\end{equation} a result in \citet{ahsanullah1988introduction}. Using the above two equations, one may verify that
\begin{equation}
	P(\Delta_k = j) = \sum_{m=0}^{j-1} \binom{j-1}{m}(-1)^m \frac{1}{(2+m)^k}.
\end{equation} Note that summing the expression over $j$ provides an alternative proof that, as reported earlier, $\mbb{E}\Delta_k=\infty$.

There are several Markov chains lurking in the background of any discussion of record values and related statistics. We list some of them that are most helpful for characterizing the behaviour of record values and record times. 

First we observe that the record counting process $\{N_n,n \geq 1\}$ is a non-stationary Markov process with transition probabilities $P(N_n=j|N_{n-1}=i)$ given by
\begin{equation}
	p_{ij} = 
	\begin{cases}
		\frac{n-1}{n}, \quad j=i\\
		\frac{1}{n}, \quad j=i+1.
	\end{cases}
\end{equation} Next note that $\{T_n,n \geq 0\}$ forms a stationary Markov chain with $P(T_0=1)=1$ and the transition probabilities $P(T_n=j|T_{n-1}=i)$ given by
\begin{equation}
	p_{ij} = \frac{i}{j(j-1)}, \quad j>i.
\end{equation} It is also obvious that $\{R_n\}$ is a Markov chain with $R_0 \sim F$ and the transitions governed by
\begin{equation}\label{eqn:conditional}
	f_{R_n|R_{n-1}} = \frac{f(r_n)}{1-F(r_{n-1})}, \quad r_n>r_{n-1}.
\end{equation} Finally, an interesting observation due to~\citet{strawderman1970law} is that $\{(R_n,\Delta_{n+1}),n \geq 0\}$ is also a Markov chain. Given the sequence $\{R_n\}$, the $\Delta_{n+1}$'s are conditionally independent geometric random variables with
\begin{equation}\label{eqn:intertimes}
	P(\Delta_{n+1}>k|\{R_n,n \geq 0\}) = P(\Delta_{n+1}>k|R_n=r_n) = (F(r_n))^k.
\end{equation} 

\subsection{Jittering the log-determinants}
As most of the analytical results for records are developed under the assumption that the random sequence is independently generated from a common continuous distribution, we would like to transform the random sequence of log-determinants, which has finite support, to its continuous counterpart. Another justification, besides that is being in accord with record values theory, is that we do not need to be concerned about ties in the random sequence. The reason is obvious: we would like to study how often strictly better approximations appear.

We will therefore transform our sequence of values with Gaussian jittering. Let $\{X_i,i \geq 1\}$ be a sequence of log-determinants generated using Algorithm~\ref{alg:SamplingSoln}. For each $X_i$, $i=1,\dots,n$, where $n$ is the sample size, we create
\begin{equation}
	Y_i | X_i \sim \text{Gaussian}(X_i,\sigma^2), \quad i=1,\dots,n,
\end{equation} so that we have a sequence of independent $\{Y_i\}$ generated from a continuous cdf. Choosing $\sigma^2$ to be very small, the values of $Y_i's$ and the behaviour of the random sequence should be identical to that of the $X_i's$ except that we have
\begin{equation}
	P(Y_l = Y_m) = 0, \quad \forall l \neq m,
\end{equation} and the corresponding random sequence of records will be strictly increasing and non-terminating.

We argue without proofs here that we can consider the $\{Y_i\}$ sequence of log-determinants with the corresponding subsets generated from a $k$-DPP distribution. More importantly, we are able to instead study any statistical behaviours of $\{Y_i\}$ and generalize them to that of $\{X_i\}$ with ignorable errors.

\subsection{Distribution of the jittered log-determinants}
\begin{figure}[!th]
	\centering
	\begin{tabular}{cc}
		\subfigure{\includegraphics[scale=0.3]{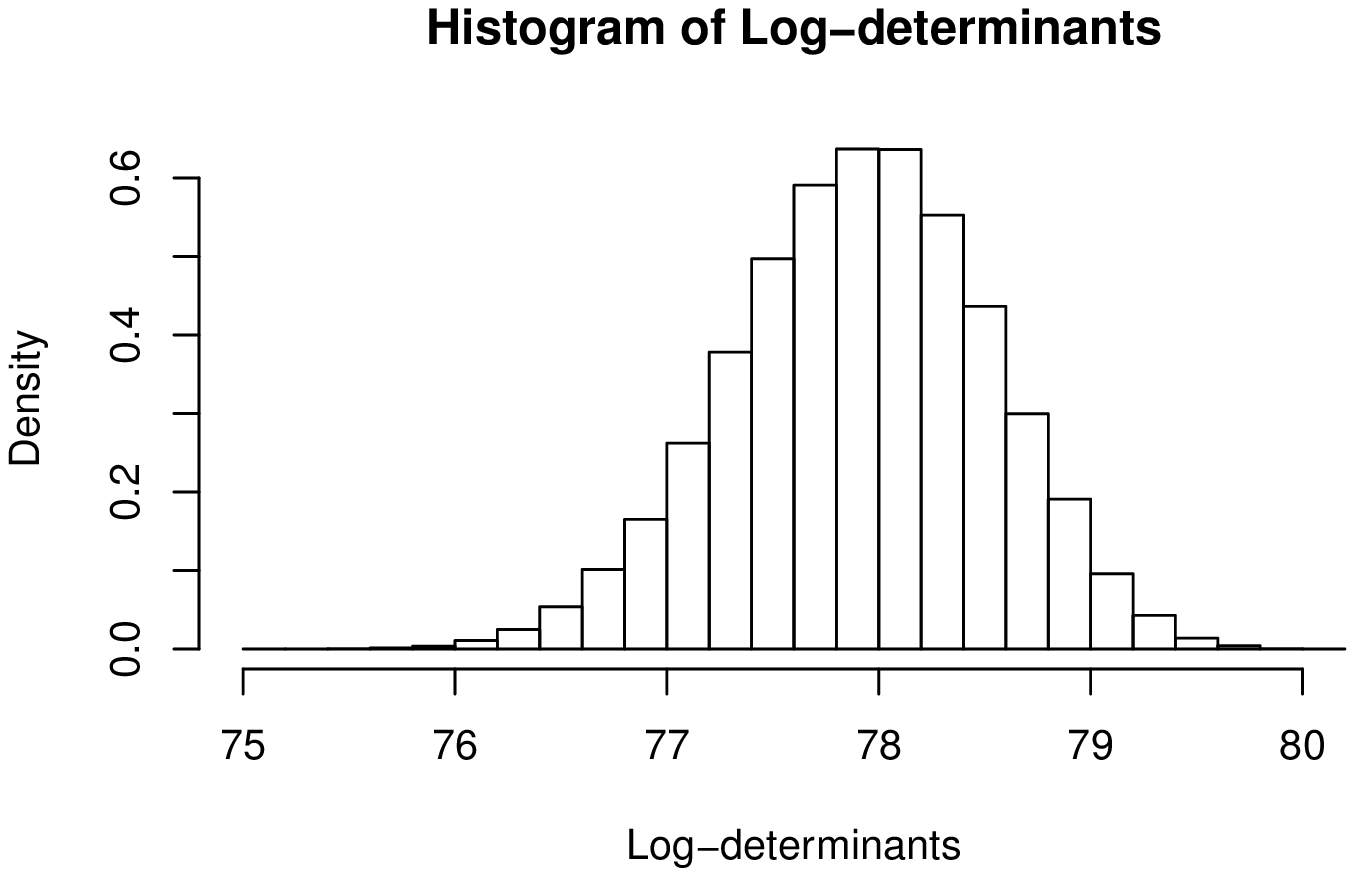}}
		\subfigure{\includegraphics[scale=0.3]{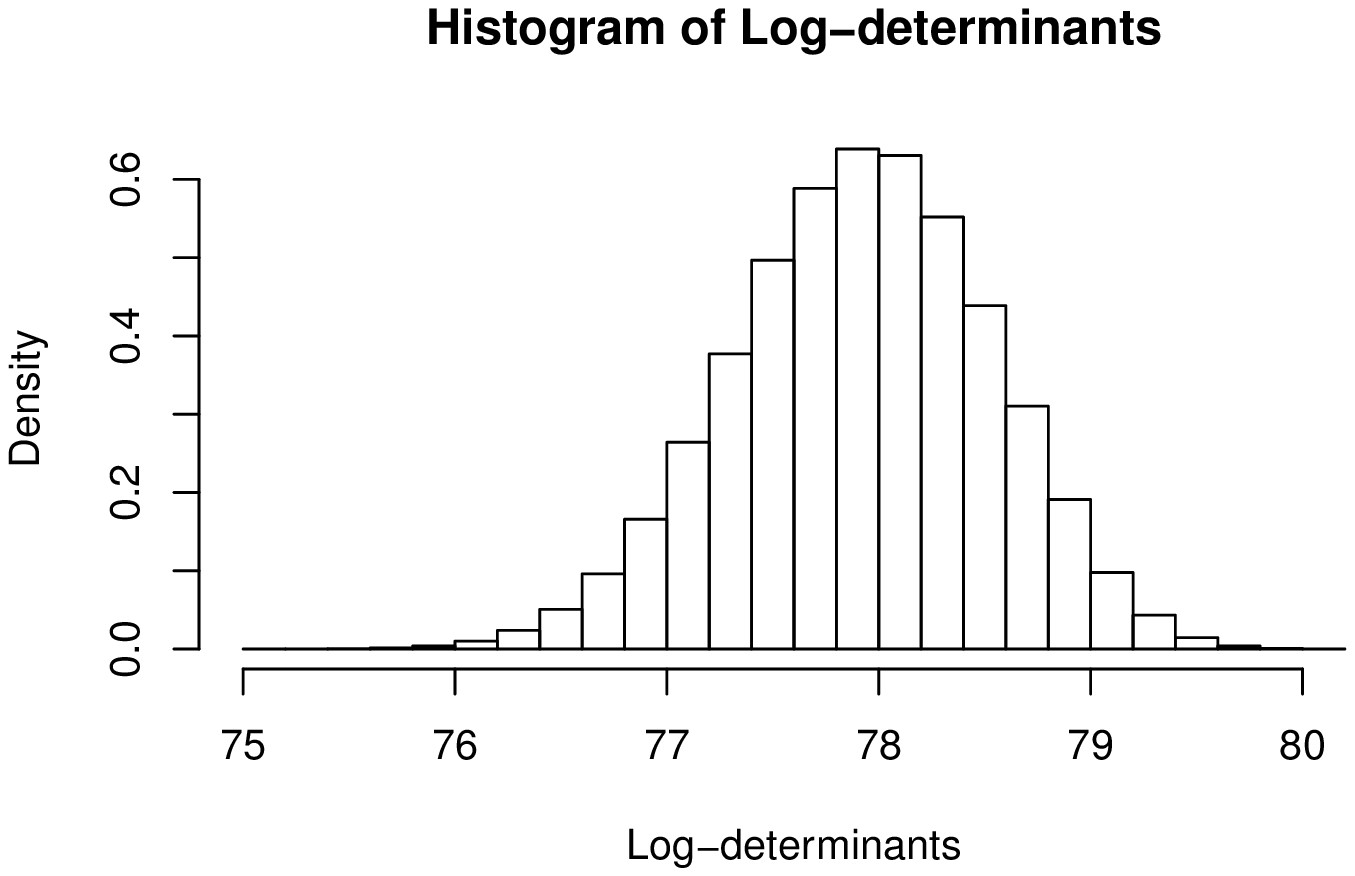}}\\
		\subfigure{\includegraphics[scale=0.3]{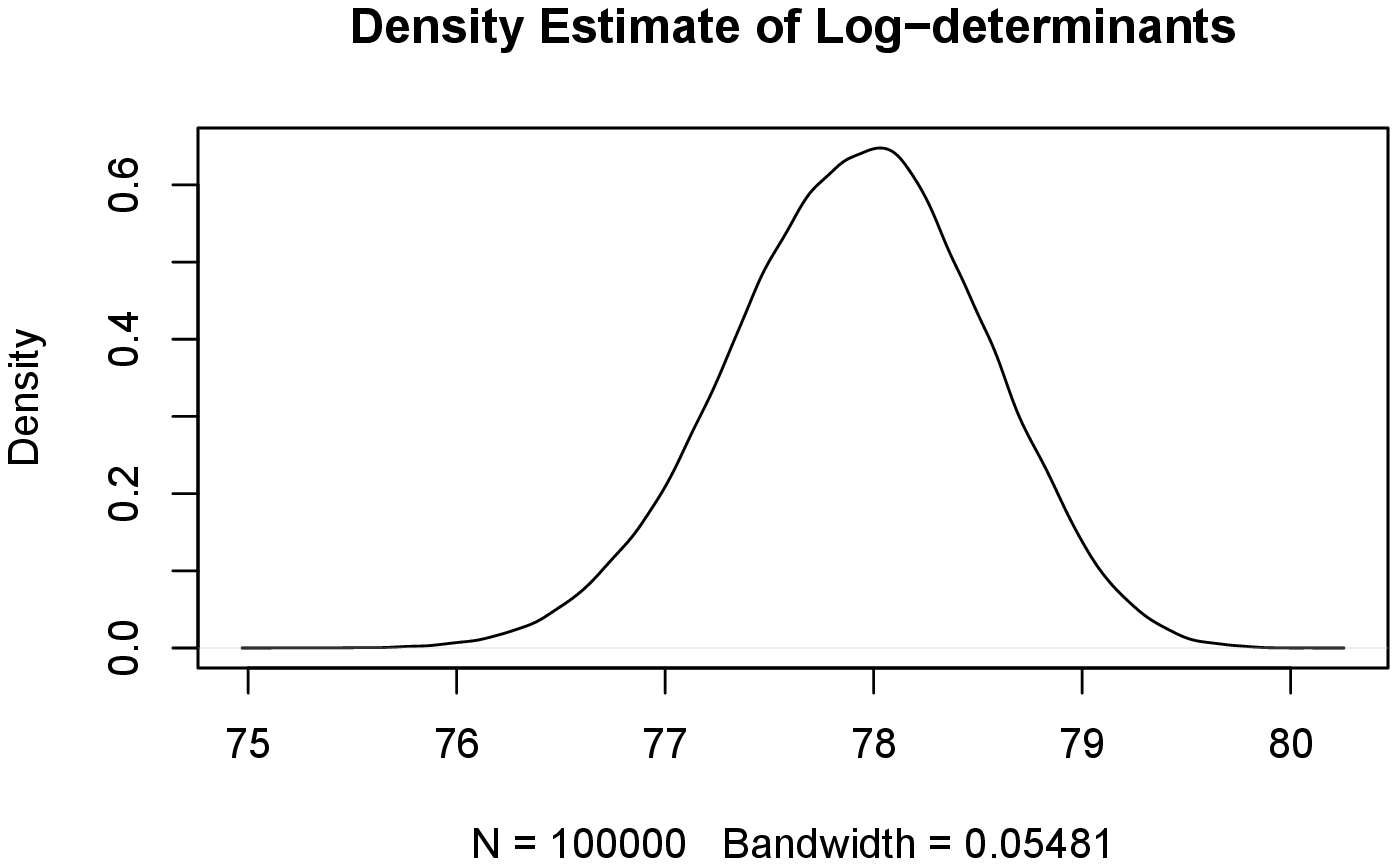}}
		\subfigure{\includegraphics[scale=0.3]{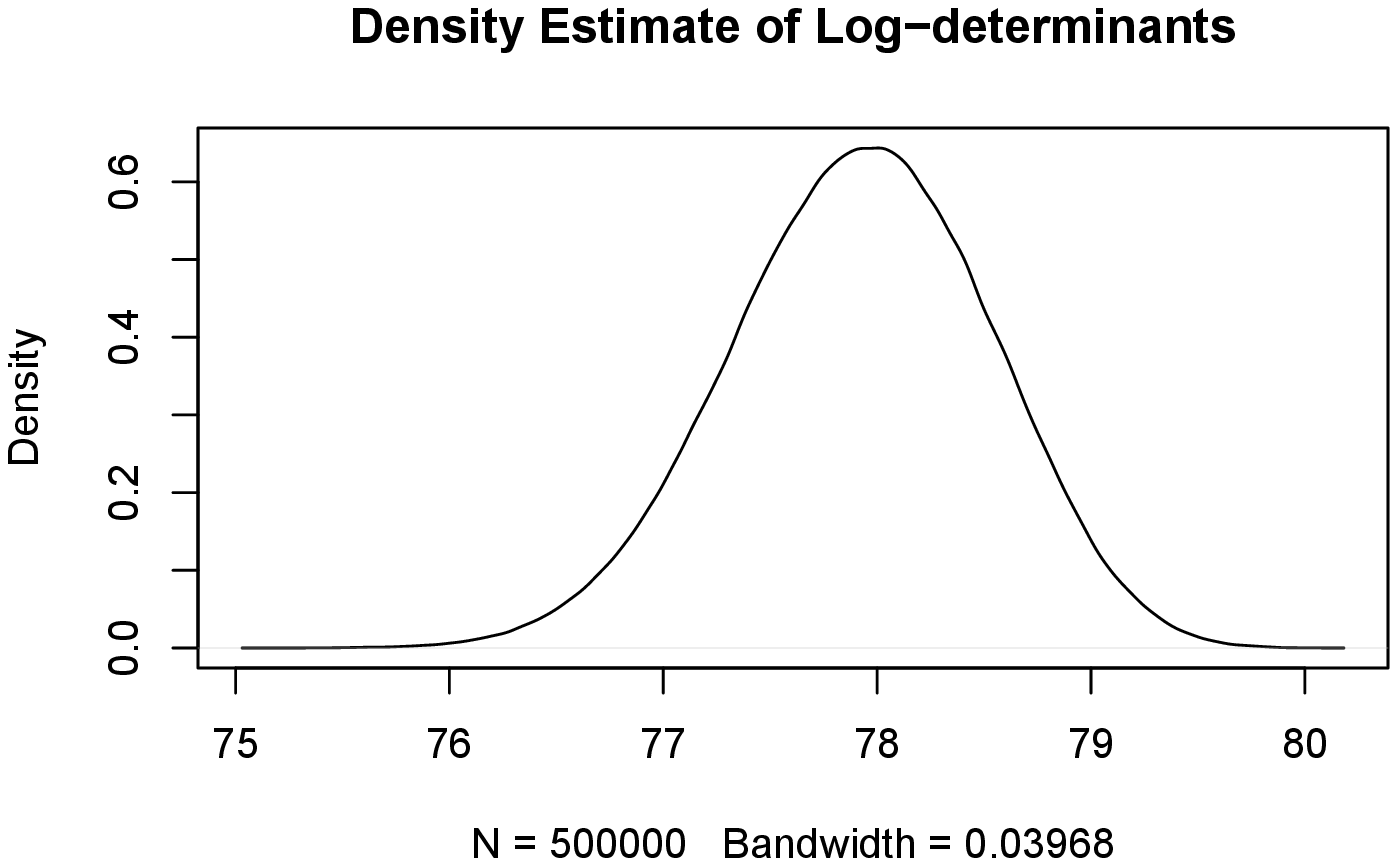}}
	\end{tabular}
	\caption{Top row: Histograms of $10$-DPP log-determinants with sample sizes $100,000$ and $500,000$ respectively; bottom row: Kernel density estimates of $k$-DPP log-determinants with sample sizes $100,000$ and $500,000$ respectively. These illustrations show the skewness of the distribution of the generated log-determinants.}\label{fig:DPPhist}
\end{figure}

The next step before applying record values theory is to find an appropriate common cdf $F$ for the jittered log-determinants. Since it would be unnecessarily complicated to study the precise analytical cdf of $Y$ due to the Gaussian jittering transformations, we fit a distribution from some parametric family that best describes the data. Figure~\ref{fig:DPPhist} shows histograms and corresponding density estimates of log-determinants generated from a $k$-DPP using the kernel described in Section~\ref{dppexp}. Note that the two plots with different sample sizes consistently convey that empirically $Y$ is distributed close to a slightly skewed Gaussian. In fact, a member from the Gaussian distribution family would fit the data well for most parts except for the tails. However, the reason that we do not consider such a distribution is that the ``right" extremes are those we concern the most. For example, in the particular kernel we used for plots, the approximations are starting getting very close to the truth after two standard deviations from the centre. For reasons such as these, we consider two distributional models to fit our data, a generalized Pareto model with a ``Peaks-Over-Threshold" fitting scheme~\citep{leadbetter1990basis}, which relies on fitting only the data exceeding a certain threshold; and an artificially left-censored Weibull model, which takes into account all the data.

\subsubsection{Extreme value theory and peaks-over-threshold}
We first provide a very brief introduction to extreme value theory and the two most common distribution families to model extreme values. Let $X_1,\dots,X_n$ be a sequence of i.i.d. random variables with common cdf $F$. Let $M_n=\max\{X_1,\dots,X_n\}$. Suppose there exists normalizing constants $a_n > 0$ and $b_n$ such that
\begin{equation}\label{eqn:EVT}
	P(\frac{M_n-b_n}{a_n} \leq y) = F^n(a_ny+b_n) \rightarrow G(y)
\end{equation} as $n \rightarrow \infty$ for all $y \in \mbb{R}$, where $G$ is a non-degenerate distribution function. According to the Extremal Types Theorem~\citep{fisher1928limiting}, $G$ must be either Fr\'{e}chet, Gumbel or negative Weibull. \citet{jenkinson1955frequency} noted that these three distributions can be merged into a single parametric family: the Generalized Extreme Value (GEV) distribution. The GEV has a distribution function defined by
\begin{equation}
	G(y) = \exp[-(1+\xi\frac{y-\mu}{\sigma})_{+}^{-1/\xi}],
\end{equation} where $(\mu,\sigma,\xi)$ are the location, scale and shape parameters respectively with $\sigma>0$. Note that $z_+=\max\{z,0\}$. The Fr\'{e}chet distribution is obtained when $\xi>0$, the negative Weibull case is obtained when $\xi<0$, and the Gumbel case is obtained when $\xi \rightarrow 0$.

From this result, \citet{pickands1975statistical} showed that the limiting distribution of normalized excesses of a threshold $\mu$ as the threshold approaches the endpoint $\mu_{\text{end}}$ of the variable of interest, is the Generalized Pareto Distribution (GPD). That is, if $X$ is a random variable which satisfies~\eqref{eqn:EVT}, then
\begin{equation}
	P(X \leq y|X > \mu) \rightarrow H(y), \quad \mu \rightarrow \mu_{\text{end}}
\end{equation} with
\begin{equation}\label{pgdp}
	H(y) = 1 - (1+\xi\frac{y-\mu}{\sigma})_{+}^{-1/\xi},
\end{equation} where again $(\mu,\sigma,\xi)$ are the location, scale and shape parameters respectively with $\sigma>0$. Note that the Exponential distribution is obtained by continuity as $\xi \rightarrow 0$.

In practice, these two asymptotical results motivated modelling block maxima with a GEV, and peaks-over-threshold with a GPD.

\subsubsection{Artificially left-censored Weibull distribution}
We also considered fitting a two-parameter Weibull distribution to all the data, except that we artificially left-censored the part of the data that is below a certain prespecified threshold. Specifically, the likelihood function for all the data is
\begin{equation}\label{LikCenWeibull}
    L(\theta) = \prod_{i=1}^{N_{nonC}} f(x_i|\theta) \prod_{j=1}^{N_{leftC}} F(x_j^{upper}|\theta)
\end{equation} with $x_i$ the $N_{nonC}$ non-censored observations, $x_j^{upper}$ the upper values defining the $N_{leftC}$ left-censored observations, and $f(x|\theta), F(x|\theta)$, the PDF and CDF of the Weibull distribution family with parameter vector $\theta$, respectively. 

Such an artificial censoring approach has been studied in the engineering literature for the computation of strength of wood-based materials. In the statistical literature, \citet{liu2018using} show that artificial censoring performs well in estimating lower quantiles.

\subsubsection{Fitting GPD and censored Weibull to jittered log-determinants}
We now use the peaks-over-threshold scheme to fit a GPD to the tail of the jittered log-determinants data. Although many estimators have been proposed in the literature (see~\citet{coles2001introduction}, for example, for a detailed discussion), for simplicity, we follow the maximum likelihood estimation procedure for point estimation for the parameters $(\mu,\sigma,\xi)$. For the censored Weibull model, we directly maximize the likelihood function in~\eqref{LikCenWeibull}.

Note that the location parameter $\mu$ for the GPD or equivalently the threshold parameter is usually not estimated in the same way as the other ones, and it is possible to estimate the other two using the MLE with a varying threshold. The main goal of threshold selection is to select enough events to reduce the variance; but not too many as we could select events coming from the central part of the distribution (i.e., not extreme events) and thus induce bias. In practice, threshold selection is usually done with the aid of exploratory tools. However, decisions are not so clear-cut for real data examples. In our study, since we have quite a lot data we set $\mu$ equal to the $90^{\text{th}}$ percentile of the log-determinants. For comparison, we artificially censor the data below the $90^{\text{th}}$ percentile in the censored Weibull model. For experiments, we have also tried the $70^{\text{th}}$ and $80^{\text{th}}$ percentiles for $\mu$ and found no material difference in the fitted distributions.

\begin{figure}[!th]
    \centering
    \includegraphics[scale=0.95]{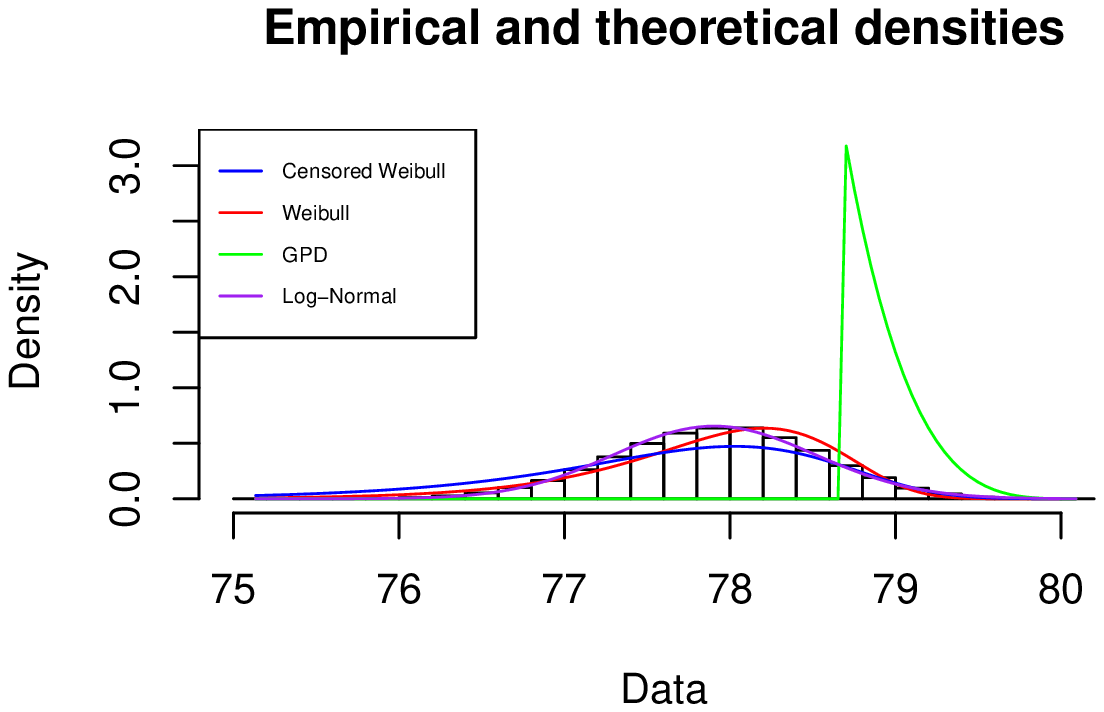}
    ~
    \centering
    \includegraphics[scale=0.95]{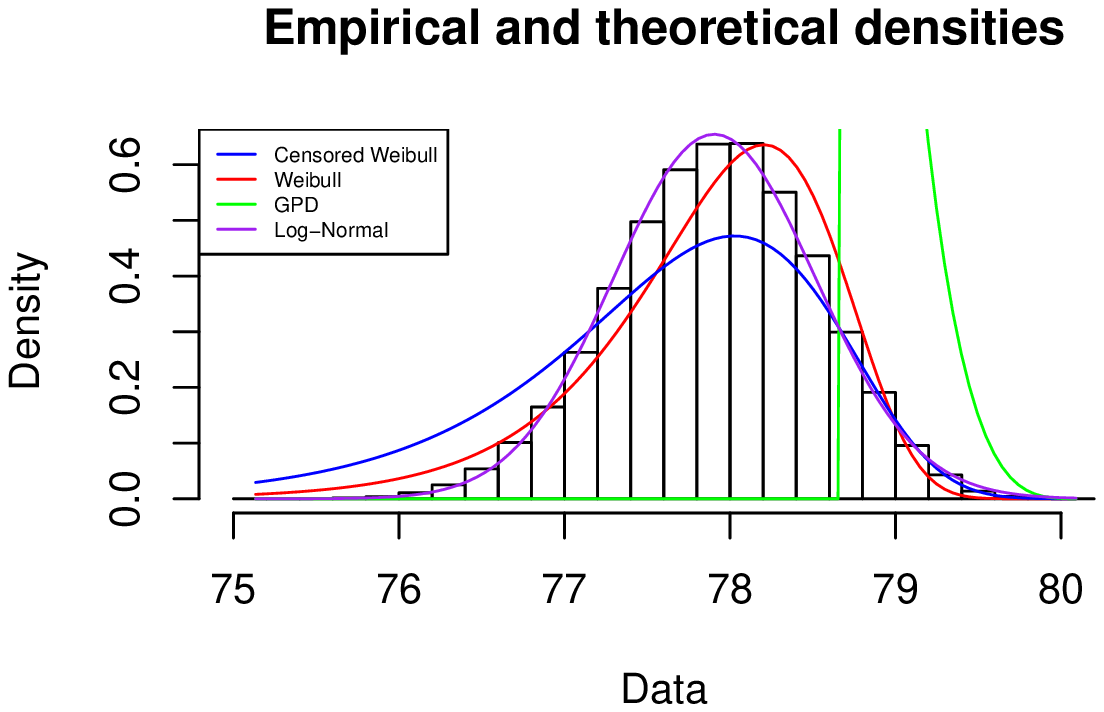}
    \caption{Full (top) and zoomed-in (bottom) versions of histograms of the log-determinants data generated from Section~\ref{dppexp} and the estimated theoretical densities. These illustrate that the GPD model allows heavy right tails while ignoring the central part of the data; and the Censored Weibull model starts perform well after the $90^{\text{th}}$ percentile threshold.}
    \label{fig:fit_densities}
\end{figure}

\begin{figure}[!th]
	\centering
	\begin{tabular}{cc}
		\subfigure{\includegraphics[scale=0.5]{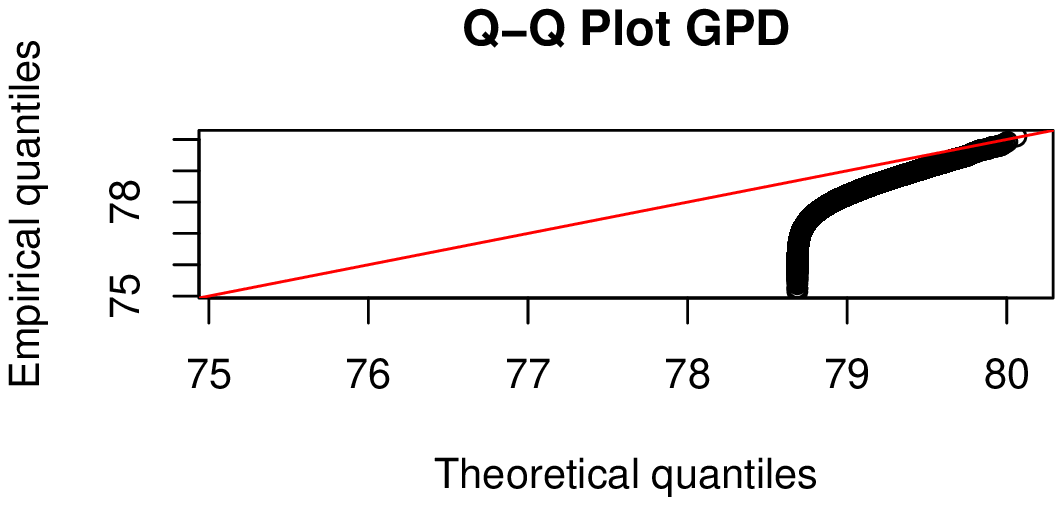}}
		\subfigure{\includegraphics[scale=0.5]{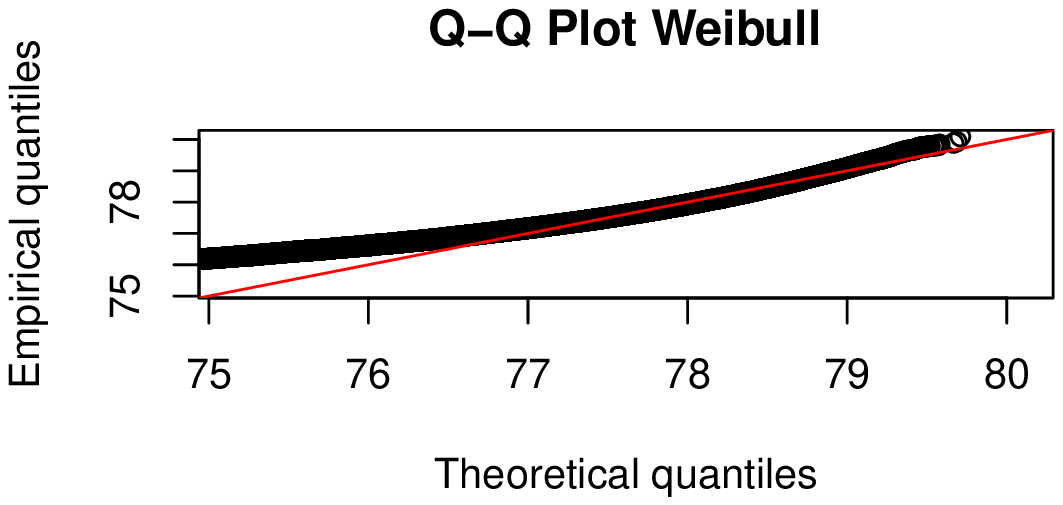}}\\
		\subfigure{\includegraphics[scale=0.5]{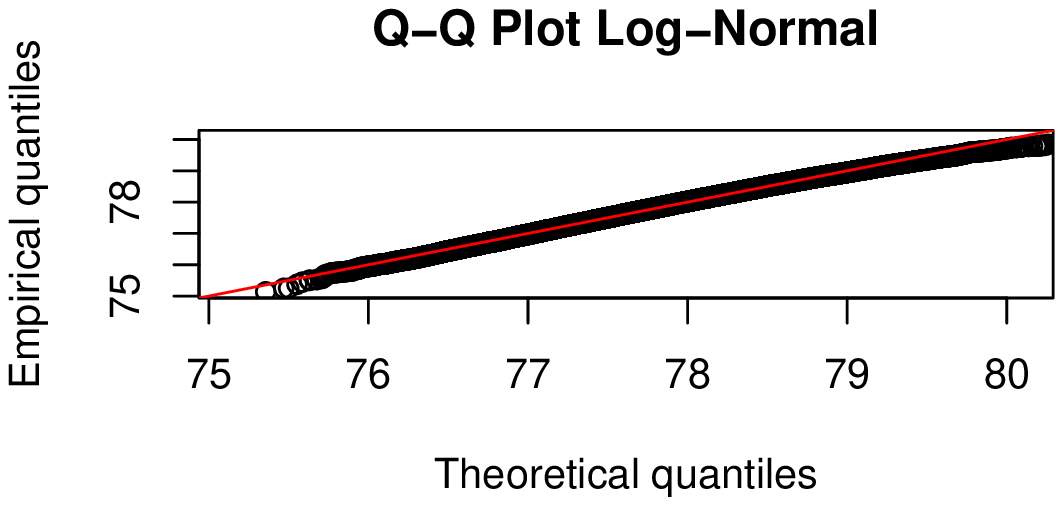}}	
		\subfigure{\includegraphics[scale=0.5]{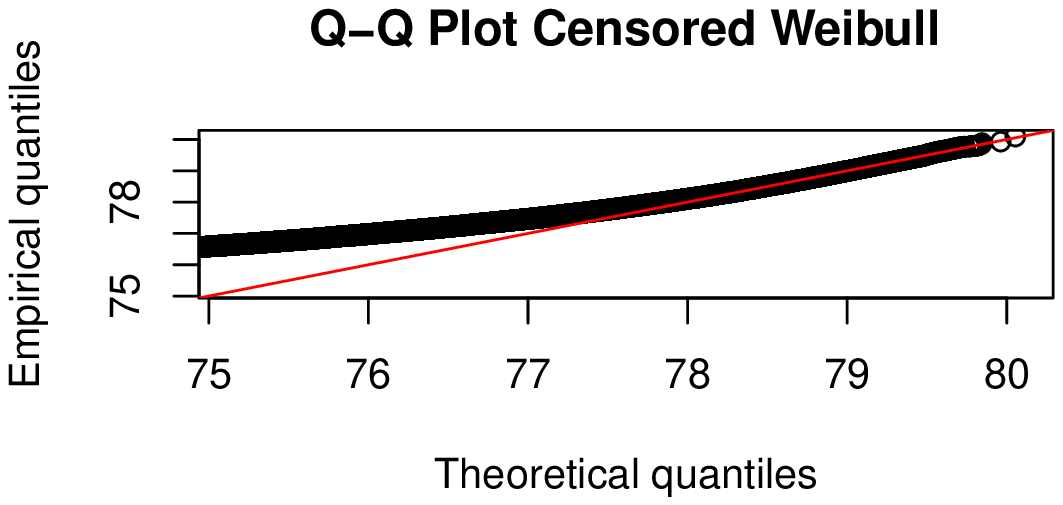}}
	\end{tabular}
	\caption{Clockwise: quantile-to-quantile plots for GPD, Weibull,censored Weibull, and Log-Normal distributions, respectively. Note that the GPD and the Censored Weibull models have superior accuracy in modeling the far-right tails of the data.}\label{fig:fit_qqplot}
\end{figure}

The results for $100,000$ log-determinants generated from the $10$-DPP with the same kernel as the one described in Section~\ref{dppexp} are shown in Figures~\ref{fig:fit_densities} and~\ref{fig:fit_qqplot}. For comparison, we also fit a Weibull and a Log-Normal distribution to the entire dataset. The analysis shows that the fitted GPD and censored Weibull are advantageous in terms of the goodness of fit in the tail part (i.e., data above $90^{\text{th}}$ percentile). Again the right tail of the data is what we are really interested in as the good approximations appear in this region. As a result, we can ignore the goodness of fit for the part of the data that is below our threshold.  

\subsection{Log-determinants records}
The end goal of studying candidate distributions for modelling jittered log-determinants is an understanding of the sequence of records generated from them. Recall from Equation~\eqref{eqn:recordsPDF} and \eqref{eqn:recordsCDF} that the distribution of records can be derived from the distribution of the original sequence. Formally, let $R_d$, $d \geq 0$ represent $d$th upper record from the jittered log-determinants sequence $\{Y_n,n \geq 1\}$. Then, we have
\begin{equation}
	f_{R_d}(r) = f(r)\frac{(-\log(1-F(r)))^d}{d!},
\end{equation}
% \begin{equation}
% 	f(y) = f_Y(y) = \frac{1}{\hat{\sigma}}(1+\hat{\xi}\frac{y-\hat{\mu}}{\hat{\sigma}})^{-1/(1+\hat{\xi})}
% \end{equation} and
% \begin{equation}
% 	F(y) = F_Y(y) = 1 - (1+\hat{\xi}\frac{y-\hat{\mu}}{\hat{\sigma}})^{-1/\hat{\xi}}.
% \end{equation} Note that the support of the last two equations is such that $y \geq \hat{\mu}$ for $\hat{\xi} \geq 0$ and $\hat{\mu} \leq y \leq \hat{\mu}-\hat{\sigma}/\hat{\xi}$ for $\hat{\xi} < 0$. Here $(\hat{\mu},\hat{\sigma},\hat{\xi})$ are MLE for $(\mu,\sigma,\xi)$.
where we can substitute $f(r), F(r)$ by any of the previously estimated models.

Before delving further into the distributional results, many distribution-free results from the record values theory can be applied to study the behaviour of our log-determinant approximations. Recall that the expected number of records in a sample of size $n$ can be approximated by $\log(n)$. In Figure~\ref{fig:dppreal} where we show the progression of $10$-DPP approximations using $100,000$ samples, the black dots represent the occurrence of records which count to $12$. Note that this is really close to the expected value, which is $\log(100,000) \approx 11.51$. Figure~\ref{fig:dppreal} tells us that the occurrence of records is more frequent at the beginning when sample sizes are small than when the sample sizes are big. This can be seen from the ``diminishing return" property of the logarithm function and it is also from an immediate application of the Markov property of the record counting process. Recall that $\{N_n,n \geq 1\}$ is a Markov process with 
\begin{equation}
	P(N_n=j|N_{n-1}=i) = 
	\begin{cases}
		\frac{n-1}{n}, \quad j=i \\
		\frac{1}{n}, \quad j=i+1.
	\end{cases}
\end{equation} Therefore, the probabilities of record-breaking in the first a few hundreds samples are much higher than those in the later samples.

\subsubsection{Conditional probabilities of record increments}
A key distributional result we use to model the behaviour of the $k$-DPP approximations is the conditional probability relating any two consecutive record values. From Equation~\eqref{eqn:conditional}, we have
\begin{equation}
	P(R_{d+1}>r|R_d=r_d) = \frac{1-F(r)}{1-F(r_d)}.
\end{equation} By letting $r=(1+\epsilon)r_d$ for some $\epsilon>0$, we have
\begin{equation}\label{eqn:condprobs}
	P(R_{d+1}>(1+\epsilon)r_d|R_d=r_d) = \frac{1-F((1+\epsilon)r_d)}{1-F(r_d)}.
\end{equation} Further, we denote the value of the above equation by $\delta_d$ for $d \geq 0$. For $\epsilon$ very small and some threshold $\delta$ very small, we can say that the probability of a small increment in the next record value is very small. In the context of $k$-DPP approximations, this is equivalent to saying that the chance of getting a better approximation is small. Although $P(R_{d+1}>r_d)=1$ by definition, the $\epsilon$ term enables us to quantify how good the next approximation can be. One other use of Equation~\eqref{eqn:condprobs}: if we have an approximation $m$ from another algorithm, then by setting $r_d=m$ the conditional probability tells us the chance that the next approximation from the $k$-DPP is better than $m$. Intuitively, this probability is an increasing function in $d$. In fact, it is possible to inversely compute $d$ such that the probability of beating $m$ is higher than some user defined threshold.

To illustrate how the conditional probabilities behave, we present numerical examples using the same sample of (jittered) log-determinants generated from the $10$-DPP in Section~\ref{dppexp}. Table~\ref{tab:condprobsGPD} and Table~\ref{tab:condprobsCensWeibull} show the occurrence of records and the associated conditional probabilities of observing better approximations (with varying $\epsilon$) given the current record values for the GPD and censored Weibull models, respectively. Note that for both models the sequences of probabilities are decreasing with the number of records with the first a few of them equal or close to one - record-breaking is more frequent at the beginning and the jumps are higher. The last record observed within this sample, though jittered, is actually the optimum value for the specific kernel. Note that the associated conditional probability is identically $0$ under the GPD model for $\epsilon = 0.001$ and $0.0005$. We also compute the conditional probabilities for the same sample but with the conditioning on the greedy approximation. The probabilities are now increasing with the progression of records - the chance of achieving better approximations (better than the greedy approximation) is higher. Note that the last record, which is better than the greedy approximation, has an associated conditional probability of $1$.  

% \begin{table}[!th]
% \begin{center}
%  \begin{tabular}{|c c c c c|} 
%  \hline
%  No. Sims. & Records & Cond. Probs. & Greedy Probs. & $\mbb{E}\Delta$ \\  
%  \hline\hline
%  1 & 78.03975 & 1.0000 & 0.0000 & 1\\ 
%  \hline
%  5 & 78.04162 & 1.0000 & 0.0000 & 1\\
%  \hline
%  7 & 78.52475 & 0.9687 & 0.0000 & 1\\
%  \hline
%  16 & 78.85338 & 0.9683 & 0.0000 & 3\\
%  \hline
%  37 & 79.36103 & 0.9417 & 0.0000 & 34\\ 
%  \hline
%  234 & 79.39688 & 0.9094 & 0.0000 & 43\\
%  \hline
%  861 & 79.66752 & 0.9085 & 0.0001 & 357\\
%  \hline
%  1758 & 79.89616 & 0.8980 & 0.0031 & 9707\\
%  \hline
%  3573 & 79.93738 & 0.8370 & 0.0408 & 26276\\
%  \hline
%  44293 & 80.09011 & 0.4937 & $>1.0000$ & $1.3386 \times 10^{10}$\\
%  \hline 
% \end{tabular}
% \end{center}\caption{$\epsilon=0.0001$}
% \end{table}

% \begin{table}[!th]
% \begin{center}
%  \begin{tabular}{|c c c c c|} 
%  \hline
%  No. Sims. & Records & Cond. Probs. & Greedy Probs. & $\mbb{E}\Delta$ \\  
%  \hline\hline
%  1 & 78.03975 & 1.0000 & 0.0000 & 1\\ 
%  \hline
%  5 & 78.04162 & 1.0000 & 0.0000 & 1\\
%  \hline
%  7 & 78.52475 & 0.9382 & 0.0000 & 1\\
%  \hline
%  16 & 78.85338 & 0.9375 & 0.0000 & 3\\
%  \hline
%  37 & 79.36103 & 0.9080 & 0.0000 & 34\\ 
%  \hline
%  234 & 79.39688 & 0.8861 & 0.0000 & 43\\
%  \hline
%  861 & 79.66752 & 0.8254 & 0.0001 & 357\\
%  \hline
%  1758 & 79.89616 & 0.8044 & 0.0031 & 9707\\
%  \hline
%  3573 & 79.93738 & 0.6957 & 0.0408 & 26276\\
%  \hline
%  44293 & 80.09011 & 0.2128 & $>1.0000$ & $1.3386 \times 10^{10}$\\
%  \hline 
% \end{tabular}
% \end{center}\caption{$\epsilon=0.0002$}
% \end{table}

\begin{sidewaystable}
% \begin{table}[!th]
\begin{center}
 \begin{tabular}{|c c c c c c c|} 
 \hline
 No. Sims. & Records & $\epsilon=0.0005$ & $\epsilon=0.001$ & $\epsilon=0.0001$ & Greedy Probs. & $\mbb{E}\Delta$ \\  
 \hline\hline
 1 & 78.03975 & 1.0000 & 1.0000 & 1.0000 &0.0000 & 1\\ 
 \hline
 5 & 78.04162 & 1.0000 & 1.0000 & 1.0000 & 0.0000 & 1\\
 \hline
 7 & 78.52475 & 0.8957 & 0.7210 & 0.9687 & 0.0000 & 1\\
 \hline
 16 & 78.85338 & 0.8691 & 0.6753 & 0.9683 & 0.0000 & 3\\
 \hline
 37 & 79.36103 & 0.7856 & 0.6053 & 0.9417 & 0.0000 & 34\\ 
 \hline
 234 & 79.39688 & 0.7756 & 0.5284 & 0.9094 & 0.0000 & 43\\
 \hline
 861 & 79.66752 & 0.6549 & 0.3518 & 0.9085 & 0.0001 & 357\\
 \hline
 1758 & 79.89616 & 0.3823 & 0.3002 & 0.8980 & 0.0031 & 9707\\
 \hline
 3573 & 79.93738 & 0.2873 & 0.1147 & 0.8370 & 0.0408 & 26276\\
 \hline
 44293 & 80.09011 & 0.0000 & 0.0000 & 0.4937 & $>1.0000$ & $1.3386 \times 10^{10}$\\
 \hline 
\end{tabular}
\end{center}
\caption{Using the GPD model: summary of occurrence of records (first two columns) and their associated conditional probabilities given the current record values (thrid through fifth columns), conditional probabilities given the greedy approximation (sixth column), and expected waiting time for the next record (last column).}\label{tab:condprobsGPD}
% \end{table}
\end{sidewaystable}

\begin{sidewaystable}
\begin{center}
 \begin{tabular}{|c c c c c c c|} 
 \hline
 No. Sims. & Records & $\epsilon=0.0001$ & $\epsilon=0.0005$ & $\epsilon=0.001$ & Greedy Probs. & $\mbb{E}\Delta$ \\  
 \hline\hline
 1 & 78.03975 & 0.9917 & 0.9585 & 0.9168 & 0.0000 & 3\\ 
 \hline
 5 & 78.04162 & 0.9857 & 0.9293 & 0.8606 & 0.0000 & 5\\
 \hline
 7 & 78.52475 & 0.9596 & 0.8478 & 0.7132 & 0.0000 & 25\\
 \hline
 16 & 78.85338 & 0.9476 & 0.8452 & 0.6498 & .0001 & 27\\
 \hline
 37 & 79.36103 & 0.9362 & 0.7600 & 0.5698 & 0.0003 & 207\\ 
 \hline
 234 & 79.39688 & 0.9154 & 0.7557 & 0.5018 & 0.0004 & 688\\
 \hline
 861 & 79.66752 & 0.9148 & 0.6380 & 0.3968 & 0.0012 & 6372\\
 \hline
 1758 & 79.89616 & 0.9124 & 0.6265 & 0.3754 & 0.0109 & 8835\\
 \hline
 3573 & 79.93738 & 0.8940 & 0.5657 & 0.3098 & 0.1130 & 11015\\
 \hline
 44293 & 80.09011 & 0.8719 & 0.4980 & 0.2384 & $>1.0000$ & 796601\\
 \hline 
\end{tabular}
\end{center}\caption{Using the censored Weibull model: summary of occurrence of records (first two columns) and their associated conditional probabilities given the current record values (thrid through fifth columns), conditional probabilities given the greedy approximation (sixth column), and expected waiting time for the next record (last column).}\label{tab:condprobsCensWeibull}
\end{sidewaystable}

\subsubsection{Conditional probabilities of inter-record times} 
Recall from Equation~\eqref{eqn:intertimes} that given the current sequence of record values, the next inter-record times are conditionally independent geometric variables with $1-p = F(r_d)$ where $p$ is the parameter for geometric distribution and $r_d$ is the $d$th record value. This distributional result gives us another way (besides the conditional probabilities of record increments) to design a stopping criteria for our $k$-DPP approximation algorithm. Specifically, we can stop sampling if the conditional probability of the waiting time for next better approximation higher than a pre-specified value is high, or the expected waiting time for the next better approximation is long. For illustration, we include in Table~\ref{tab:condprobsGPD} and Table~\ref{tab:condprobsCensWeibull} the expected waiting times given the current sequence of record values. As anticipated, the last expected waiting time is huge, which indicates  a stopping point for the algorithm. In fact, as discussed before, the last record value is already the theoretical maximum and hence should be the stopping point. 

Essentially, a combination of the conditional probabilities and the expected waiting times provides us an incisive tool to analyze our $k$-DPP approximations. In the numerical example above, these values tell us that we have reached an approximation such that the probability of obtaining a better solution is very low and the waiting time for this hypothetical better solution is expected to be very long, that is, we should stop sampling.

\section{Concluding remarks}\label{conclusion}
In this paper we have studied the problem of maximum-entropy design of spatial monitoring networks. In particular, we have examined a stochastic search algorithm based on the DPP for approximating optimal designs; we explored properties of record values and their related statistics and applied them to model jittered log-determinants (approximations) generated by the DPP-based algorithm. We obtained interesting results based only on simple distributional facts (i.e. distribution and density functions). In particular, many of the observed behaviours seen in the $ k $--DPP approximations, such as the quick escalation at the beginning and the diminishing return near the end, can be explained by record value theory. We also developed informative stopping rules for the search algorithm using a combination of distributional results for record values and inter-record times. 

In principle our approach leads to the the optimal, model based design if enough iterations are allowed. But it's not clear that this should be the ultimate goal. After all, all models are wrong.  So an alternative we plan to explore is that of stopping early to obtain a model--guided randomized design. This randomized design option would provide some robustness over strictly model based designs.  And it could also be seen as having an advantage over purely probability based designs like stratified sampling, where hard boundaries are drawn to subdivide the $ N $ possibilities to force  the sample to be well distributed over the population.  The DPP provides an alternative route to the objective of a well--distributed sample, albeit without the artificially hard boundaries imposed in stratified sampling. 

We have not presented the wide range of possible objectives that could be accommodated by the DPP approach. For instance, instead of the entropy, the DPP kernel could be an intersite distance matrix, thereby forcing a geographical diversity in the selected design points. Following the idea in the previous paragraph, we could thus obtain a randomized version of the space--filling design. Alternatively the kernel could take the intersite differences in  say historical annual site response averages.  
% Or the kernel could be a positively weighted  sum of the various kernels above as a compromise among the different design objectives. Still another option would be to treat these various objectives as constituting a  multi--attribute design selection problem.  These two approaches will be compared in future work.

In studying record values, we have made several choices in terms of statistical models and parameters. For instance, we computed the conditional probabilities of the next $1+\epsilon$ better records with prespecified values of $\epsilon$'s. The choice of $\epsilon$ is intuitively connected to the desired gain in information from the corresponding new design of the network. This connection will be explored in future work to rigorously quantify the relationship between $\epsilon $ and the information gain. 
	
	In fitting the GPD and censored Weibull distributions we have hand-picked the thresholds. We plan to explore systematic ways of estimating the thresholds from the data. In order for the log-determinants sampled from the $ k $-DPP to fit into the classical record model, we introduced the idea of Gaussian jittering. Although it worked quite well in our examples, we plan to rigorously quantify the approximation errors resulting from the jittering.

{\bf Acknowledgements.}  We are indebted to Professor Alexander Bouchard--C\^{o}t\'{e} for pointing out the possible link between the DPP process and spatial design. The work was partially supported by Natural Sciences and Engineering Research Council of Canada. The numerical examples were enabled in part by support provided by WestGrid and Compute Canada Calcul Canada.

% \section*{References}
% \begin{thebibliography}{99}
% \end{thebibliography}
\bibliographystyle{apalike}
\bibliography{ref}

\end{document}